\documentclass[reprint, secnumarabic,amssymb, nobibnotes, aps, dvipsnames]{revtex4-1}

 \usepackage[dvipsnames]{xcolor}
 \usepackage{amsmath}
 \usepackage[title]{appendix}
 \usepackage{array}
 \usepackage{calc}
 \usepackage{float}
 \usepackage{graphicx}
 \usepackage{hyperref}
 \usepackage{longtable}
 \usepackage{multirow}
 \usepackage{tikz}
 
\usetikzlibrary{calc, matrix, positioning, shapes, shapes.geometric}
\newcommand{\diff}[1]{\mathrm{d}#1}

\DeclareMathOperator{\atantwo}{arctan2}
\newcommand{\red}[1]{{{#1}}}

\begin{document}

	\title{Superfluid Vortices in Four Spatial Dimensions}
	\author{Ben McCanna}
	\email[]{bdm375@student.bham.ac.uk}
	\affiliation{School of Physics and Astronomy, University of Birmingham, Edgbaston, Birmingham B15 2TT, UK}
	\author{Hannah M. Price}
	\affiliation{School of Physics and Astronomy, University of Birmingham, Edgbaston, Birmingham B15 2TT, UK}
	\begin{abstract}
        Quantum vortices in superfluids have been an important research area for many decades. 
        Naturally, research on this topic has focused on two and three-dimensional superfluids, in which vortex cores form points and lines, respectively. Very recently, however, there has been growing interest in the quantum simulation of systems with four spatial dimensions; this raises the question of how vortices would behave in a higher-dimensional superfluid. In this paper, we begin to establish the phenomenology of vortices in 4D superfluids under rotation, where the vortex core can form a plane. In 4D, the most generic type of rotation is a ``double rotation" with two angles (or frequencies). We show, by solving the Gross-Pitaesvkii equation, that the simplest case of equal-frequency double rotation can stabilise a pair of vortex planes intersecting at a point. This opens up a wide number of future research topics, including \red{into realistic experimental models; unequal-frequency double rotations; the stability and potential reconnection dynamics of intersecting vortex surfaces;} and the possibility of closed vortex surfaces.
	\end{abstract}
	\maketitle

\begin{figure}
    \includegraphics[scale=0.91]{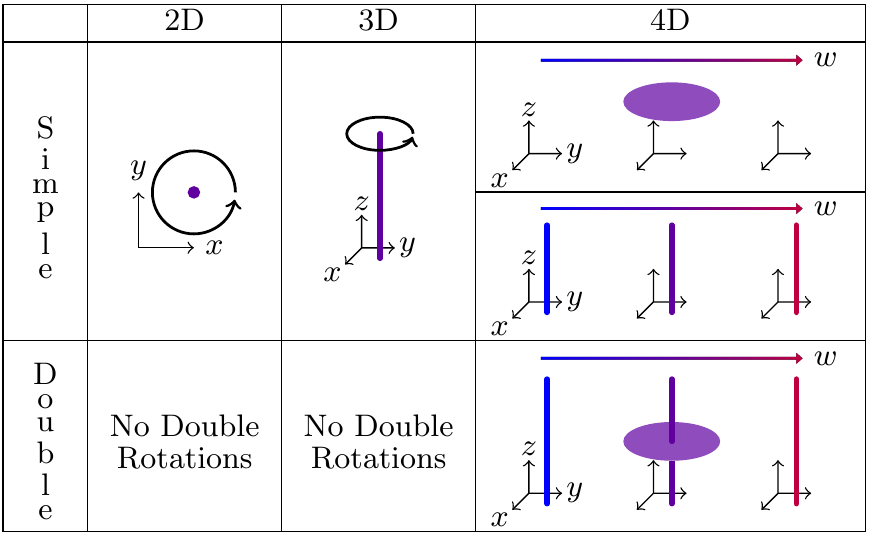}
    \caption{Sketch of minimal vortex structures, stabilised for different system dimensionalities (\textit{columns}) and types of rotation (\textit{rows}). Here, ``simple" and ``double" indicate rotations with one or two planes of rotation respectively, as discussed in the text.
    In 2D and 3D, only simple rotations exist, stabilising vortex cores as a point and line, respectively, about which the superfluid rotates (black arrow). In 4D space (shown as 3D cross-sections coloured according to \(w\) value), both types of rotation exist, leading to a richer vortex phenomenology. In 4D, equal-frequency double rotations can lead to a new type of vortex configuration consisting of two vortex planes intersecting at a point, while simple rotations stabilise a single vortex plane. In these sketches, a vortex plane appears either as a line persisting for all \(w\) (lines of varying colour), or as a plane for a particular \(w\) value (purple disc), depending on the rotation plane. Note that in the 4D column we have omitted the arrow indicating superfluid motion.}
    \label{fig:cartoon}
\end{figure}

Quantum vortices are fundamental topological excitations of superfluids, which have been widely studied for many years~\cite{pitaevskii2003, cooper2008rapidly, fetter2009, madison2000, madison2001, matthews1999,abo-shaeer2001}. Unlike a lot of many-body phenomena, vortices can be understood at the mean-field level through the Gross Pitaevskii equation (GPE)~\cite{pitaevskii2003}. A superfluid vortex consists of a local density depletion within the ``vortex core", around which the superfluid circulates. In 2D and 3D superfluids, this vortex core forms a point and a line respectively, as sketched in Fig~\ref{fig:cartoon}. Vortices have an associated energy cost, but can be stabilised by rotation of the superfluid~\cite{fetter2009,cooper2008rapidly}, or equivalently by artificial magnetic fields~\cite{dalibard2011colloquium,Cooper_2019,Ozawa2019Photonics}.

While research has so far naturally focused on vortices in 2D and 3D superfluids, there is growing interest in simulating systems with four spatial dimensions. This is thanks to experimental and theoretical investigations of 4D physics in topological pumping~\cite{Kraus_2013,Lohse_2018,Zilberberg_2018}, high-dimensional parameter spaces~\cite{sugawa2018second,lu2018topological, kolodrubetz2016measuring, wang2020exceptional} and electric circuits with high connectivity~\cite{wang2020circuit,Price2018, yu2019genuine,li2019emergence,ezawa2019electric}, as well as proposals for engineering 4D systems using ``synthetic dimensions"~\cite{Price_2015,Ozawa2016}. The latter, in particular, opens up the prospect of being able to explore higher-dimensional superfluids with artificial gauge fields. In this approach, ``synthetic dimensions" are built by coupling together the internal states of cold atoms~\cite{Boada2012,Celi2014,Mancini2015,Stuhl2015,gadway2015atom, An_2017,Price2017,salerno2019quantized, viebahn2019matter,barbiero2019bose, chalopin2020exploring}, photonic systems~\cite{Ozawa2016,Yuan2016, ozawa2017synthetic, lustig2019photonic, Yuan2018,Yuan2019, yuan2020creating, Dutt2020} and other platforms~\cite{baum2018setting, price2019synthetic,crowley2019half,boyers2020exploring}. Such degrees of freedom are then reinterpreted as lattice coordinates in a new direction, increasing the effective system dimensionality, while providing straightforward ways to realise artificial magnetic fields~\cite{ozawa2019topological}, and hence mimic rotation in a higher-dimensional space.

The potential of synthetic dimensions for reaching 4D with (for example) ultracold bosonic atoms~\cite{Price_2015, viebahn2019matter} motivates the question of how superfluid vortices behave in higher dimensions. \red{In this paper, we take an initial step in this direction by exploring the 4D GPE under rotation, with local atom-atom interactions. This is chosen as a minimal model, which naturally extends a standard textbook problem to 4D in order to establish basic aspects of 4D vortex physics. More realistic models for experiments will depend on the specific synthetic-dimension implementation chosen, and are likely to include other effects, such as lattices and unusual interactions with respect to the synthetic dimension, that will further enrich the possible vortex states, but will go beyond the current work. We also note that while our main motivation for studying the 4D GPE is as an initial stepping-stone towards possible synthetic-dimension experiments, this model is also plausible as a description of low-temperature interacting bosons in a hypothetical 4D universe (see Appendix~\ref{App:4DGPE}, and~\cite{Wodkiewicz,Stampfer,Trang}), and so is of mathematical interest for generalising classic results about superfluid vortices to higher dimensions.
}

To investigate vortices in 4D, we must first appreciate that rotations (or equally, magnetic fields) in higher dimensions can have a fundamentally different form; all rotations in two and three dimensions are so-called ``simple rotations", while in 4D, generic rotations are ``double rotations"~\cite{lounesto2001}. This difference will be discussed in more detail later, but can be understood in brief by noting that in 2D/3D every rotation has a single rotation plane and angle, while in 4D there can be two independent rotation planes, e.g. the \(xy\) and \(zw\) planes, each with their own angle of rotation.

In this paper, we show that equal-frequency double rotation of a 4D superfluid can stabilise a vortex structure formed by two vortex planes intersecting at a point, while a simple rotation stabilises a single vortex plane, as sketched in Figure \ref{fig:cartoon}. We obtain our results, firstly by using a phase ansatz to numerically solve an effective 2D radial equation, and secondly by numerically solving the full 4D GPE under rotation. This generalisation of superfluid vortices to higher dimensions opens up many avenues of future research, such as questions concerning the unequal-frequency case; reconnections of vortex planes; possible curvature of vortex surfaces; and more realistic setups capturing experimental details.

\section{Review of vortices in 2D and 3D superfluids}
\label{sec:VortexReview}

We begin by reviewing the basic properties of 2D and 3D vortices, in order to lay the groundwork for our discussion of 4D superfluids. We consider systems of weakly-interacting bosons described by a complex order parameter, \(\psi\), which obeys the time-independent GPE with no external potential~\cite{pitaevskii2003}
\begin{equation}
	-\frac{\hbar^2}{2m}\nabla^2\psi + g|\psi|^2\psi = \mu\psi,
	\label{eq:GPE}
\end{equation}
where \(m\) is the particle mass, \(g\) is the interaction strength, and \(\mu\) is the chemical potential. A hydrodynamic description can be obtained from this equation by substituting \(\psi=\sqrt{\rho}e^{iS}\), where \(\rho\) is the superfluid density, and \(S\) is the phase~\cite{pitaevskii2003}. The velocity field, \(\mathbf{v}=\frac{\hbar}{m}\nabla S\), is irrotational wherever \(S\) is well behaved. A consequence of this property is that a superfluid supports quantized vortices. This can be seen by noting that the superfluid circulation around a closed loop \(C\) is quantised as
\begin{equation}
    \oint_C \mathbf{v}\cdot\diff\mathbf r = \frac{\hbar}{m} \left[\Delta S\right]_{C}, 
\end{equation}
where \([\Delta S]_C\) is the phase winding~\cite{fetter2009}. Since \(\psi\) is single-valued, we must have \([\Delta S]_C = 2\pi k\), where \(k\) is the integer winding number (or vortex charge)~\cite{pitaevskii2003}. Smoothly deforming the loop cannot change \(k\) as long as vortices are avoided. This can only be true if \(\mathbf{v}\) diverges like \(1/r\) as the distance \(r\) from a vortex core goes to zero. Since particles cannot have infinite velocity \(\rho\) must vanish in this same limit. The region of density depletion is known as the vortex core; in 2D, this is localised around a point, and in 3D around a line, as shown in Fig~\ref{fig:cartoon}. More generally, vortices must be localised in two directions.

As is well known, the density profile around the vortex core can be calculated directly by applying the GPE to a homogenous superfluid with a single vortex~\cite{pitaevskii2003}. By defining the uniform background density \(n\), the healing length \(\xi\) can be introduced, which satisfies \(\hbar^2/m\xi^2 = gn = \mu\)~\footnote{Note that many authors include the factor of \(1/2\) in the definition of \(\xi\).}, and which physically is the distance over which \(\rho\) typically varies. Hereafter, we rescale \(\mathbf{r} \to \xi \mathbf{r}\), and \(\psi \to \sqrt{n} \psi\) such that Eq~\eqref{eq:GPE} becomes dimensionless as
\begin{equation}
    -\frac{1}{2}\nabla^2\psi + |\psi|^2\psi = \psi
    \label{eq:DGPE}.
\end{equation}
A rotationally symmetric vortex state in 2D has the form \(\psi = f_k(r)e^{ik\theta}\), where \((r,\theta)\) are polar coordinates centred on the vortex core, \(f_k(r)\) is real, and \(k\) is the winding number. Substituting this into Eq~\ref{eq:DGPE} gives~\cite{pitaevskii2003}
\begin{equation}
	-\frac{1}{2}\left( \Delta_r - \frac{k^2}{r^2} \right)f_k + f_k^3 - f_k = 0,
	\label{eq:f2D}
\end{equation}
where \(\Delta_r = \partial^2/\partial r^2 + (1/r)\partial/\partial r\). This equation has no closed-form solution, but does admit the asymptotic forms \(f_k(r) = O(r^{|k|})\) as \(r \to 0\), and \(f_k(r) = 1 - O(r^{-2})\) as \(r \to \infty\)~\cite{fetter2009}. The crossover between these two behaviours occurs at around the healing length. Note that a straight vortex line in an otherwise homogeneous and isotropic \(3D\) superfluid has this same profile, with \((r,\theta)\) defined in the plane perpendicular to the vortex line~\cite{pitaevskii2003}.

Using this density profile the energy cost of a vortex relative to the ground state can be evaluated. For a singly charged vortex (\(k=1\)) the energy can be written as
\begin{equation}
    E_1(R) = \mu N \left(\frac{\xi}{R}\right)^2\ln\left(2.07\frac{R}{\xi}\right),
    \label{eq:VEnergy}
\end{equation}
where \(N\) is number of bosons, and \(R\) is the radius of the superfluid in the plane orthogonal to the vortex core. Eq~\eqref{eq:VEnergy} is valid in any number of dimensions. Vortices can be energetically stabilised by rotation (or equivalently an artificial magnetic field), whereby Eq~\eqref{eq:GPE} is modified in 3D by adding the term \(-\mathbf{\Omega}\cdot\mathbf{L}\psi\) to the left hand side, with \(\mathbf{L} = -i\hbar\mathbf{r}\times\nabla\) the angular momentum operator, and \(\mathbf{\Omega}\) the frequency vector~\cite{pitaevskii2003}. This term reduces the energy of a state containing a vortex aligned with the rotation, making it more energetically favourable.

\section{Simple and double rotations}
\label{sec:4DRotations}

Given the intrinsic link between rotation and vortices, we will now discuss the different types of rotations possible in 4D, as compared to lower dimensions, in preparation for our discussion of vortices in 4D superfluids below. 

In three dimensions or fewer, every rotation is ``simple"; this means that the rotation is specified by a rotation angle \(\alpha\in(-\pi,\pi]\), and a plane of rotation which is unique up to translation. Under rotation, the points on the plane of rotation remain on the plane, but are displaced through the angle \(\alpha\). Generalising to \(D\) dimensional space, simple rotations have \(D-2\) eigenvectors with eigenvalue one, all of which are orthogonal to every vector in the rotation plane. For example, a rotation about the \(z\) axis in 3D has the \(xy\) plane (defined by \(z=0\)) as its rotation plane, and fixes any point along the \(z\) axis. We may write this as a matrix in the standard basis as
\begin{equation}
    \begin{pmatrix}
        \cos{\alpha} & -\sin{\alpha} & 0 \\
        \sin{\alpha} & \cos{\alpha} & 0 \\
        0 & 0 & 1
    \end{pmatrix}.
\end{equation}
We can think of this as a rotation of 2D space (spanned by \(x\) and \(y\)) extended into a third (\(z\)) direction. Similarly, simple rotations in 4D can be thought of as rotations of 3D space extended into a fourth direction. Labelling the fourth axis as \(w\), our previous example becomes a rotation about the \(zw\) plane (defined by \(x=y=0\)), given in matrix form by
\begin{equation}
    \begin{pmatrix}
        R(\alpha) & 0 \\
        0 & I
    \end{pmatrix}, \ \text{where} \
    R(\alpha) = 
    \begin{pmatrix}
        \cos{\alpha} & -\sin{\alpha} \\
        \sin{\alpha} & \cos{\alpha} \\
    \end{pmatrix},
    \label{eq:4Dsimple}
\end{equation}
and \(I\) is the 2D identity. Note that there are six Cartesian coordinate planes in 4D, so the rotation group \(SO(4)\) has six generators, and the representation of these generators (which physically describe angular momentum) as spatial vectors no longer works in 4D as it does in 3D. The set of fixed points of a simple rotation in 4D are a plane, not a line, and this fixed plane is completely orthogonal to the plane of rotation, by which we mean that every vector in one plane is orthogonal to every vector in the other.

In contrast to 2D and 3D, in four dimensions, we can also have ``double rotations", which generically have only one fixed point, and two completely orthogonal planes of rotation each with a corresponding rotation angle~\cite{lounesto2001}. To visualise this, consider a double rotation in the \(xy\) and \(zw\) planes represented by the matrix \footnote{any double rotation can be brought into this form by an orthogonal transform}
\begin{equation}
    \begin{pmatrix}
        R(\alpha) & 0 \\
        0 & R(\beta)
    \end{pmatrix},
    \label{eq:DoubleRot}
\end{equation}
for angles \(\alpha,\beta\in(-\pi,\pi]\). For those familiar with certain 4D quantum Hall models, this is analogous to generating a second Chern number by applying magnetic fields in two completely orthogonal planes \cite{Price_2015,Ozawa2016,Lohse_2018,Zilberberg_2018,Mochol_Grzelak_2018}. Double rotations are in fact the generic case of rotations in 4D, as if either \(\alpha\) or \(\beta = 0\), the rotation reduces to the special case of simple rotation discussed above~\cite{lounesto2001}. From here on we will refer to the two planes of rotation as planes 1 and 2 respectively and focus only on so-called ``isoclinic" double rotations for which \(\alpha=\beta\). 

\red{Before continuing, it is worth noting that isoclinic rotations have an additional symmetry. To see this, we remember that, as introduced above, generic double rotations have one fixed point and two planes of rotation, with corresponding angles \(\alpha, \beta \in (-\pi,\pi]\). Vectors in \(\mathbb{R}^4\) which do not lie in these rotation planes are displaced through an angle between \(\alpha\) and \(\beta\)~\cite{lounesto2001}. However, if \(\alpha=\beta\), then this means that every vector is displaced by the same angle. As a consequence, for a given isoclinic rotation there is a continuum of pairs of completely orthogonal planes that can each be though of as the two planes of rotation. In other words, isoclinic rotations therefore no longer have two unique planes of rotation, although they still have a single fixed point. However, numerically we break this degeneracy since the phase winding of our initial state picks out the \(xy\) and \(zw\) planes in particular. We can also anticipate that a more experimental model would likely break this symmetry too, e.g. through the inclusion of lattices or through inherent differences between real and ``synthetic" spatial dimensions. }

\section{Vortex planes in 4D}
\label{sec:VortexPlanes}

Now that we have discussed some of the geometry of rotations in 4D we are ready to study the associated vortex physics. As above, we consider a superfluid described by the GPE in the absence of external potentials, but now with atoms free to move in four spatial dimensions.

\begin{figure}
    \centering
    \begin{tikzpicture}
        \def\dx{0.48\linewidth}
    	\node at (0,0) {\includegraphics[width=\dx]{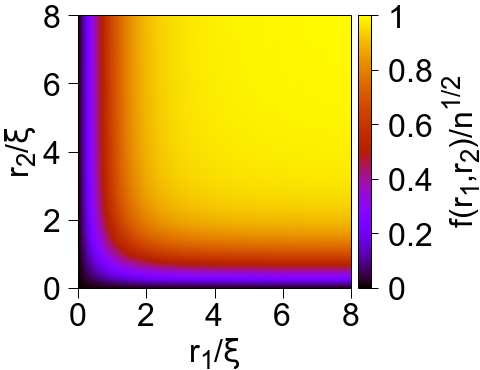}};
    	\node at (-0.2\linewidth,0.2\linewidth) {(a)};
    	\node at (\dx,0) {\includegraphics[width=\dx]{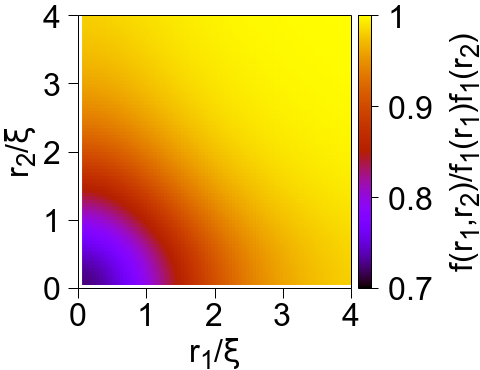}};
    	\node at (0.27\linewidth,0.2\linewidth) {(b)};
    \end{tikzpicture}
	\caption{(a) Numerical solution of Eq~\eqref{eq:f} for \(f(r_1,r_2)\), with \(k_1=k_2=1\), showing the density profile for an intersecting pair of vortex planes in 4D, as a function of the two polar radii. (b) The ratio of the solution in (a) to the product approximation \(f_1(r_1)f_1(r_2)\), where \(f_j(r_j)\) is the well-known 2D vortex profile governed by Eq~\ref{eq:f2D}. This shows that the product approximation works well away from the intersection as expected, but fails in a small region around \(r_1=r_2=0\). Numerical \red{parameters and details are given in the main text}.}
	\label{fig:radial}
\end{figure}

The simplest case to consider is that of a 4D superfluid under a constant simple rotation. As shown in Eq~\eqref{eq:4Dsimple}, a simple rotation can be viewed as a 3D rotation extended into a fourth dimension, hence stabilising a vortex plane, as sketched in Fig~\ref{fig:cartoon}. The corresponding order parameter profile is \(\psi = f_k(r_1)e^{ik\theta_1}\), where \((r_1,\theta_1)\) are plane polar coordinates in the plane of rotation, and \(f_k(r)\) is the solution of Eq~\eqref{eq:f2D}. As this is independent of the other two coordinates, the vortex core becomes a plane; this is directly analogous to the extension of a point vortex in 2D into a line in 3D. We have verified this result numerically, as shown in \red{Appendix~\ref{App:Simple}}. This can be understood as the natural extension of vortices into 4D, as the extra dimension plays no role, and the vortex plane is \red{homotopically} characterised by a \(\mathbb{Z}\) topological winding number, as in 2D and 3D. \red{For a more detailed discussion of homotopy classification of vortex planes in 4D, see Appendix~\ref{App:Homotopy}.}

In contrast we expect that double rotation, being an intrinsically 4D (or higher) phenomenon, will lead to more interesting vortex configurations. To address this problem, we look for the ground states of the 4D GPE in a doubly rotating frame
\begin{equation}
	\left[-\frac{\hbar^2}{2m}\nabla^2 + g|\psi|^2 - \Omega_1L_1 - \Omega_2L_2\right] \psi = \mu\psi,
	\label{eq:GPER}
\end{equation}
where \(\Omega_j\) and \(L_j\) are the rotation frequency and angular momentum operator in plane \(j\). In Cartesian coordinates (\(x,y,z,w\)), \(L_1 = -i\hbar(x\partial_y - y\partial_x)\), and \(L_2 = -i\hbar(z\partial_w - w\partial_z)\). For simplicity we will adopt double polar coordinates \((r_1,\theta_1,r_2,\theta_2),\) defined by
\begin{equation}
    (x,y,z,w) = (r_1\cos\theta_1,r_1\sin\theta_1,r_2\cos\theta_2,r_2\sin\theta_2), \nonumber
\end{equation}
such that \(L_j = -i\hbar\partial_{\theta_j}\). The simple rotation case discussed before corresponds to \(\Omega_2=0\), where the vortex core spans plane 2. In this paper we focus on equal-frequency doubly rotating superfluids, that is \(\Omega\equiv\Omega_1=\Omega_2\).

The fact that \(L_1\) and \(L_2\) generate a double rotation means that they commute. We may look for a solution which is a simultaneous eigenstate of both angular momentum operators; therefore we propose an ansatz for the ground state under rotation of the form
\begin{equation}
    \psi(\mathbf{r}) = f(r_1,r_2)e^{ik_1\theta_1 + ik_2\theta_2},
    \label{eq:DoubleAnzatz}
\end{equation}
where \(f(r_1,r_2)\) is real and the \(k_{j}\) are integer phase winding numbers in each rotation plane. This phase profile corresponds to the superfluid circulating in both planes simultaneously, about both vortex cores. We have suppressed the dependence of \(f\) on each \(k_j\) for brevity, and in all numerical results both winding numbers are one. This state exhibits a phase singularity when either \(r_j=0\), so we require \(f(0,r_2)=f(r_1,0)=0\) from the same reasoning as in 2D and 3D. In other words, this describes a pair of completely orthogonal vortex planes that intersect at a single point as illustrated in Fig~\ref{fig:cartoon}, and which are characterised by \(\mathbb{Z}\times \mathbb{Z}\) topological winding numbers \red{(see Appendix~\ref{App:Homotopy})}. Intersection of two planes at a point is only possible in 4D or higher and, in fact, is the generic case in 4D. This is in contrast with 3D, where \red{the intersection of lines is a special case}, and so vortex lines intersect and reconnect at specific times~\cite{koplik1993,nazarenko2003analytical,zuccher2012,allen2014}.

To examine our ansatz, we now proceed to numerically solve for the density profile, under this phase constraint.
Substituting the ansatz [Eq~\eqref{eq:DoubleAnzatz}] into the GPE [Eq~\eqref{eq:GPE}] in 4D, and de-dimensionalising in the same way as in the 2D case, we obtain the following equation for \(f(r_1,r_2)\)
\begin{equation}
	-\frac{1}{2}\left( \Delta_{r_1} - \frac{k_1^2}{r_1^2} + \Delta_{r_2} - \frac{k_2^2}{r_2^2} \right)f + f^3 - f = 0,
	\label{eq:f}
\end{equation}
where \(\Delta_{r_j} = \partial^2/\partial r_j^2 + (1/r_j)\partial/\partial r_j\).
Since each vortex produces only a local density depletion, we expect that \(f(r_1,r_2) \sim f_{k_2}(r_2) \ \text{as} \ r_1 \to \infty\) and equally for \((1 \leftrightarrow 2)\), where \(f_k(r)\) is the point vortex solution of Eq~\eqref{eq:f2D}. \red{Note that this limiting ``boundary condition" can be satisfied by a product, \(f_{k_1}(r_1)f_{k_2}(r_2)\), of 2D density profiles in each plane. However, this form fails to solve the full equation due to the non-linear \(f^3\) term. This product form therefore gives a natural approximation to compare to, and we expect it to fail significantly only in the vicinity of the origin, where both \(f_{k_j}(r_j)\) differ appreciably from unity.}

To verify this, \red{and find the full density profile,} we have solved Eq~\eqref{eq:f} by imaginary time evolution \red{within a discretised grid in \((r_1\), \(r_2)\) space with hard-wall boundary conditions at a radius \(R=100\xi\) in each plane (\(r_j=R\)), and at the origin in each plane (\(r_j=0\)). The latter condition is required due to the centrifugal term diverging at the vortex cores; consequently the precise location of the vortex cores was an assumption in these calculations. We used a forward Euler time-discretization and second order finite differences in space. We chose a large value of \(R\) compared to \(\xi\) so that we could examine the vortex cores within a homogeneous region. (Future studies could include the effect of additional trapping potentials, such as harmonic traps along some or all directions.) We were able to achieve a resolution of \(0.05\xi\), and the calculations were converged until the relative change in chemical potential and particle number over one timestep converged below \(10^{-14}\).}

The results for \(k_1=k_2=1\) are shown in Fig~\ref{fig:radial}(a), where we observe the expected local density depletion around the vortex cores when either \(r_1=0\) or \(r_2=0\). We also compare our numerical solution with the product approximation, \(f_{1}(r_1)f_{2}(r_2)\), in Fig~\ref{fig:radial}(b); we observe that the product approximation is very accurate \red{except within a distance of roughly} \(\gtrsim\xi\) from the intersection point, as expected. Immediately around the intersection, the product approximation fails, overestimating the density by a factor of about \(4/3\).

Just as in the 2D case we can use our calculation of the density profile to find the energy of this vortex configuration relative to the state with no vortices. Defining independent radii \(R_j\) in each plane, such that \(r_j \leq R_j\), \red{we find numerically (see Appendix~\ref{App:Energy})} that the energy is approximately given as
\begin{equation}
    E_{k_1,k_2}(R_1,R_2) = E_{k_1}(R_1) + E_{k_2}(R_2),
    \label{eq:EnergySum}
\end{equation}
where \(E_k(R)\) is the single-vortex energy given in Eq~\eqref{eq:VEnergy}. This can be understood from the superfluid kinetic energy \(\int \rho v^2 \diff^4 r\), which is the main contribution to the energy of a vortex. The velocity field is given by \(\mathbf{v} = \mathbf{v}_1 + \mathbf{v}_2\) where \(\mathbf{v}_j = \frac{k_j}{r_j}\hat{\theta}_j\) is the velocity induced by vortex \(j\). As \(\mathbf{v}_j\) lies in plane \(j\), we see that \(\mathbf{v}_1\cdot\mathbf{v}_2 = 0\) and so the hydrodynamic vortex-vortex interaction term, \(\int \rho\mathbf{v}_1\cdot\mathbf{v}_2 \diff^4r\), vanishes. The total kinetic energy integral therefore splits into a sum of the individual kinetic energies. Note that this argument relies on the assumptions that the two vortex cores have no curvature and are completely orthogonal to each other.

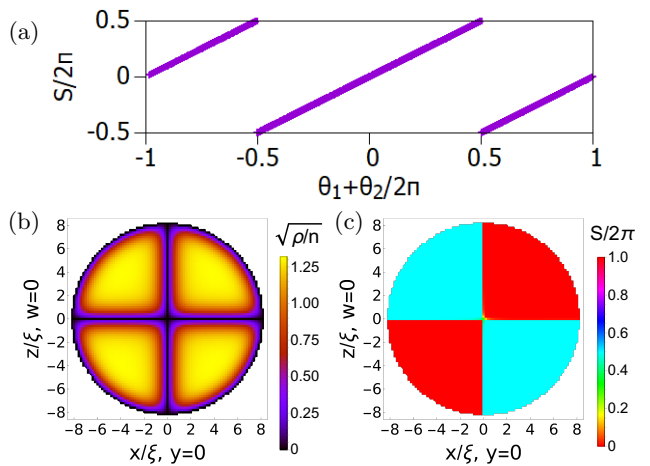
\begin{figure}
    \centering
    \begin{tikzpicture}
        \def\height{90pt};
        \def\width{115pt};
        \def\offset{8pt};
        
        \node at (-\width,1.3*\height) {(a)};
        \node at (0,\height)
        {\includegraphics[width=0.85\linewidth]{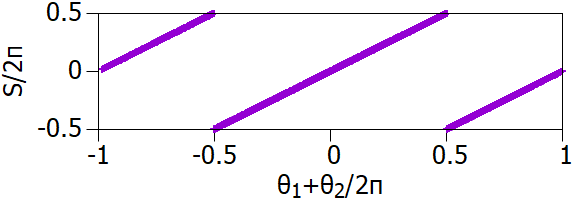}};
        
        \node at(-\width,.5*\height) {(b)};
        \node at (0,0)
        {\includegraphics[width=0.95\linewidth]{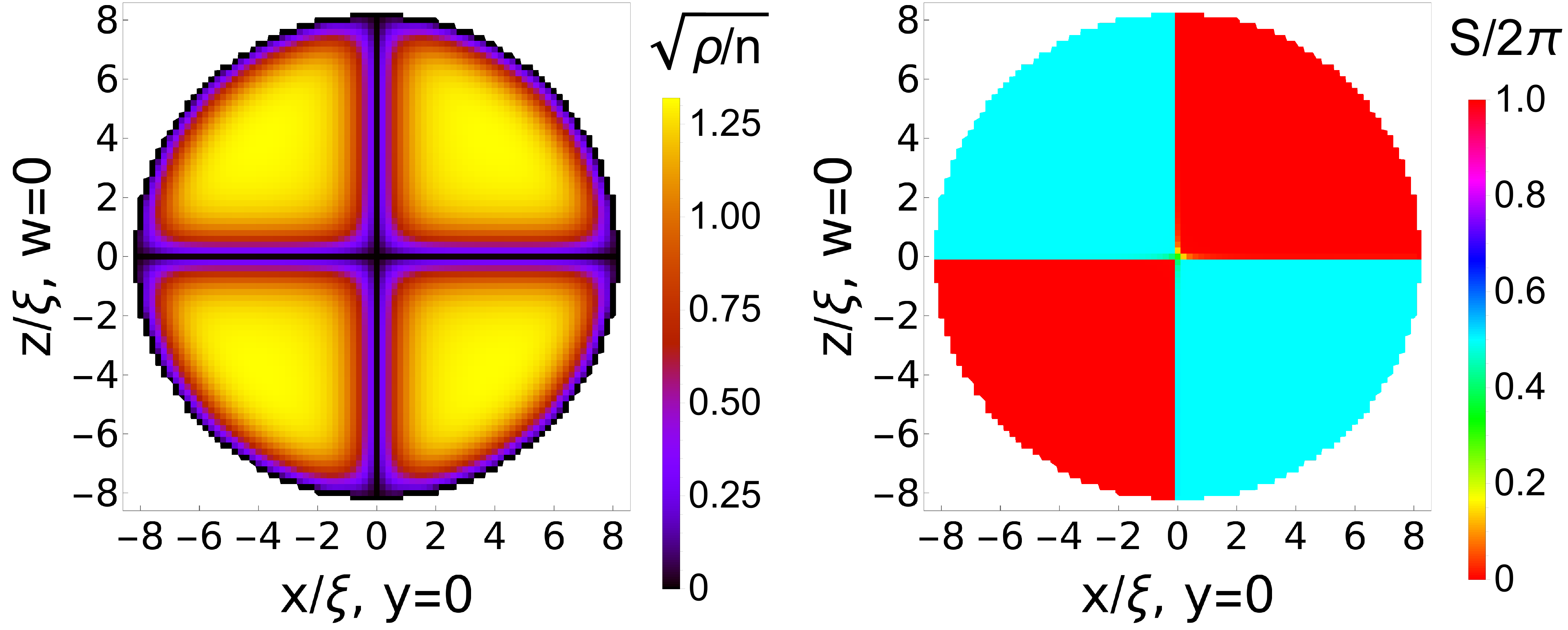}};
        \node at (\offset,.5*\height) {(c)};
    \end{tikzpicture}
    \caption{Numerical results from imaginary time evolution of the doubly-rotating 4D GPE in a ball geometry of radius \(\sim 8\xi\), given an initial state with phase profile \(\theta_1 + \theta_2\) and additional noise. (a) The phase of the final state at each point within the 4D ball vs the sum of the two polar angles, showing perfect agreement with the phase profile of our ansatz [Eq~\eqref{eq:DoubleAnzatz}]. The density (b) and phase (c) profiles of the final state for the 2D slice in which \(y=w=0\); these are consistent with our ansatz, as well as the density profile shown in Fig~\ref{fig:radial}. We can interpret this final state as containing two vortex planes, one at \(x=y=0\), and one at \(z=w=0\). \red{Further 2D cuts of this state are given in Appendix~\ref{App:Double}}.}
    \label{fig:4DNumerics}
\end{figure}

In order to confirm the existence and stability of the intersecting vortex plane state we have performed imaginary time evolution with the 4D GPE under \red{both simple and} double rotation [Eq~\eqref{eq:GPER}] directly on a 4D Cartesian grid \red{within a 4D ball of radius \(R=8.25\xi\) with a hard-wall boundary. A hyper-sphere rather than a hyper-cube was chosen as the majority of the 4D volume of a hyper-cube is taken up by regions "in the corners", that is, outside of the hyper-sphere that just fits inside.} This allowed us to relax our above constraint on the phase profile, at the cost of smaller numerical system sizes. \red{Again, we used the forward Euler method for time-discretization and second order finite differences in space.  We were able to obtain resolutions of up to \(0.2\xi\), and by repeating simulations at different resolutions, we checked that our main conclusions were qualitatively insensitive to the coarse-graining of the numerics. At the system sizes and resolutions we have been able to reach, the homogeneous region extends over a few healing lengths. The calculations were converged to an accuracy threshold of \(10^{-12}\).

A benefit of performing calculations with all four coordinates is that we were able to test our ansatz by allowing the phase to evolve, and by removing the boundary condition at \(r_j=0\) mentioned previously. More precisely, we used an initial state with homogeneous density away from the edge of the ball, and a phase profile given by \(\atantwo(y,x)+\atantwo(w,z)\), for the doubly rotating case, and \(\atantwo(y,x)\) for the singly rotating case. We tested the robustness of our results to noise (up to 20\% of the background value) added to the real and imaginary parts of the initial \(\psi\). Note that we measure the applied frequency in units of the critical frequency of a single vortex in a homogeneous 2D disk of the same radius as our 4D ball; this is given (in our units) by~\cite{pethick2002}
\begin{equation}
    \Omega_{\text{crit}}^{2D} = \mu \log(2.07 R/\xi) \left( \frac{R}{\xi}\right)^2. 
\end{equation}
For the results shown in Fig~\ref{fig:4DNumerics} both the frequencies of rotation used were roughly \(2.5\Omega_{\text{crit}}^{2D}\). Further work could investigate the effect of double rotation with unequal frequencies.
}

For a suitable range of frequencies \(\Omega\) we find good agreement between the stationary state obtained from the full 4D numerics and our ansatz for two intersecting vortex planes, as shown in Fig~\ref{fig:4DNumerics}. Panel (a) shows that the phase profile of the state after relaxation perfectly agrees with that of the ansatz. Panels (b) and (c) show the density and phase profiles, respectively, for the 2D cut in which \(y=w=0\). As can be seen the density drops to zero along the lines \(x=0\) and \(z=0\), corresponding to the intersections of each vortex core with the plane of the cut, as expected. Further two dimensional cuts of this state are given in \red{Appendix~\ref{App:Double}}.

\section{Discussion and Conclusions}
\label{sec:Conclusions}

In this paper, we have shown that the simple rotation of an idealised 4D superfluid can stabilise a vortex plane, while equal-frequency double rotations can lead to two vortex planes intersecting at a point which do not interact hydrodynamically. This significantly extends the phenomenology of superfluid vortices, demonstrating that new effects can emerge in higher spatial dimensions even within mean-field theory.

\red{It is important to note that we have studied an idealised model, which allows us to explore vortex physics in 4D without experimental details that depend on how the synthetic dimension is implemented~\cite{Boada2012,Celi2014,Mancini2015,Stuhl2015,gadway2015atom,An_2017,Price2017,salerno2019quantized,chalopin2020exploring}. The main differences between our work and possible experiments are, firstly, that the majority of practical implementations would lead to (tight-binding) lattice models, whereas we have considered four continuous dimensions as a theoretical first step. Adding a lattice should introduce rich additional effects particularly when the lattice spacing is comparable to or greater than other length scales. However when this spacing is very small, it should be possible to approximate a lattice model with a continuum model in the mean-field regime as we have considered here. Furthermore, synthetic-dimension schemes can include unusual effects, which are very dependent on the specific experimental implementation. In terms of the tight binding description previously mentioned, these complications can include position-dependent hopping strengths, limited numbers of sites, and long-range interactions~\cite{Boada2012,Celi2014,Mancini2015,Stuhl2015,gadway2015atom,An_2017,Price2017,chalopin2020exploring}. For the sake of generality as well as simplicity we have therefore chosen an idealised model, which can then be adapted in different ways for promising experimental scenarios in further work.

We also note that Eq~\eqref{eq:GPE} has \(SO(4)\) (4D rotational) symmetry, which would be broken in any experiment due to inequivalence of the synthetic and real spatial dimensions. Numerically, we break this symmetry with the phase anzatz, which was assumed in the radial case, and imposed on the initial state in the Cartesian case. However, we do still assume an \(SO(2)\) symmetry in each of the \(xy\) and \(zw\) planes to obtain the effectively 2D radial equation [Eq~\eqref{eq:f}], and simplify the corresponding numerics. In the Cartesian case, we also chose a boundary condition (a hard-wall at some radius from the origin) that preserved these in-plane symmetries. In synthetic dimension experiments, on the other hand, the most common boundary condition is an open boundary condition which is independent of the other dimensions~\cite{Boada2012,Celi2014,Mancini2015,Stuhl2015,gadway2015atom,An_2017,Price2017,salerno2019quantized,chalopin2020exploring}. Hence a more experimentally relevant geometry would involve one or more dimensions which have their own independent hard-wall boundary conditions - for example a ``spherinder" boundary specified by \(\{\mathbf{r}\in\mathbb{R}^4|x^2+y^2+z^2=R^2\}\cup\{\mathbf{r}\in\mathbb{R}^4|w = \pm L\}\) for some \(R\) and \(L\). Investigating the effect of breaking one or both of these in-plane rotational symmetries geometrically is an interesting and natural next step for future work. 
}

\red{As well as a first step towards understanding future experimental models, this work also opens up many interesting theoretical research directions.} Natural next steps include the study of 4D superfluids doubly rotating at unequal frequencies, and 4D generalisations of previously studied questions from 2D and 3D~\cite{fetter2009,cooper2008rapidly}. Firstly, closed vortex surfaces in 4D would naturally generalise the vortex loops that arise in 3D~\cite{pitaevskii2003}, but with potentially an even richer classification when non-orientability and surfaces of higher genus are included~\cite{gallier2013}. Secondly, vortex lines in 3D are known to \red{dynamically} reconnect upon intersection~\cite{koplik1993,nazarenko2003analytical,zuccher2012,allen2014}, whereas here we have shown that \red{completely orthogonal} intersecting vortex planes in 4D \red{form a stationary state} stabilised by rotation. It is an open question whether \red{vortex planes reconnect if they are not completely orthogonal, and this question could have relevance to the general case of unequal-frequency double rotation. For example, intuitively, we would expect an adiabatic change from \(\Omega_2=\Omega_1\) to \(\Omega_2>\Omega_1\) would cause the vortex in plane 2 (inducing rotation in plane 1) to tilt towards plane 1 to benefit from the now larger rotational energy discount in plane 2. Finally, in} the longer term this work opens up questions related to the inclusion of strong interactions and the 4D fractional quantum Hall effect, as well as the study of models with more interesting order parameter spaces~\cite{Kawaguchi, Machon}, potentially hosting non-Abelian vortices. \\

{\it Acknowledgements:} We thank Tomoki Ozawa, Mike Gunn, Iacopo Carusotto, Mark Dennis and Russell Bisset for helpful discussions. 
This work is supported by the Royal Society via grants UF160112, RGF\textbackslash{}EA\textbackslash{}180121 and RGF\textbackslash{}R1\textbackslash{}180071, as well as by EPSRC.

\begin{appendices}
\renewcommand\appendixname{Appendix}
    \setcounter{equation}{0}
    \setcounter{figure}{0}
    \setcounter{table}{0}
    \renewcommand{\theequation}{A\arabic{equation}}
    \renewcommand{\thefigure}{A\arabic{figure}}
    \newlength{\picwidth}
    
    \section{The 4D GPE for an Idealised 4D Bosonic Gas}
    \label{App:4DGPE}
    
    The 4D GPE is a natural and mathematically simple generalisation of the 3D GPE, allowing for easy comparison to superfluid vortex physics in lower dimensions. In this section, we also point out that the 4D GPE can be motivated as the proper description of interacting bosons in a hypothetical 4D universe, and so is an interesting theoretical model in its own right. As is well known, the use of the GPE to describe a system of interacting bosons relies on taking the Hartree-Fock approximation and replacing the interaction potential by a contact (Dirac delta) potential. The latter trick is in turn justified by looking at the low energy limit of the solutions for two-particle scattering. In this limit, the solutions are spherically symmetric (s-wave) and correspond to solutions for a contact interaction with the same scattering length as the original potential. While this argument is usually applied only in three dimensions and below, it has also been generalised to arbitrary dimensions~\cite{Wodkiewicz,Stampfer,Trang}, showing that the dimensionality only affects the contact interaction strength, and the form of the short-range singularities that must be removed from the scattering equation. The interaction strength can be considered arbitrary due to scale invariance of the GPE in the absence of an external potential, and the singularities have no effect on the GPE. Hence, it can be concluded that the GPE should be a valid mean-field description of interacting bosons at low energy in 4D. 
    
    \section{Additional Numerical Results}
    
    \subsection{Simple Rotations}
    \label{App:Simple}
    
    As described in \red{Section~\ref{sec:VortexPlanes}}, we expect that a simple rotation should be able to stabilise a single vortex plane, extending the concept of 2D point vortices and 3D line vortices straightforwardly to four-dimensional systems. Assuming the rotation is in plane 1 (as defined in Section~\ref{sec:4DRotations}), this would correspond to a condensate wavefunction of the form: 
    
    \begin{eqnarray}
    \psi=f(r_1, r_2) e^{i k_1 \theta_1} \label{eq:single}
    \end{eqnarray}
    
    with \(f(0,r_2)=0\), and such that this wave-function approximately takes the form \(\psi\propto (x+iy)\) near the vortex core. 
    
\begin{figure*}
            \centering
            \begin{tikzpicture}
                \setlength{\picwidth}{0.49\linewidth}
                \newlength{\dx}
                \setlength{\dx}{0.25\picwidth}
                \def\dy{1.85};
                \coordinate (letter) at (-1.5*\dx,.75*\dy);
                \node at (letter) {(a)};
                \node at (0,0) {\includegraphics[width=\picwidth]{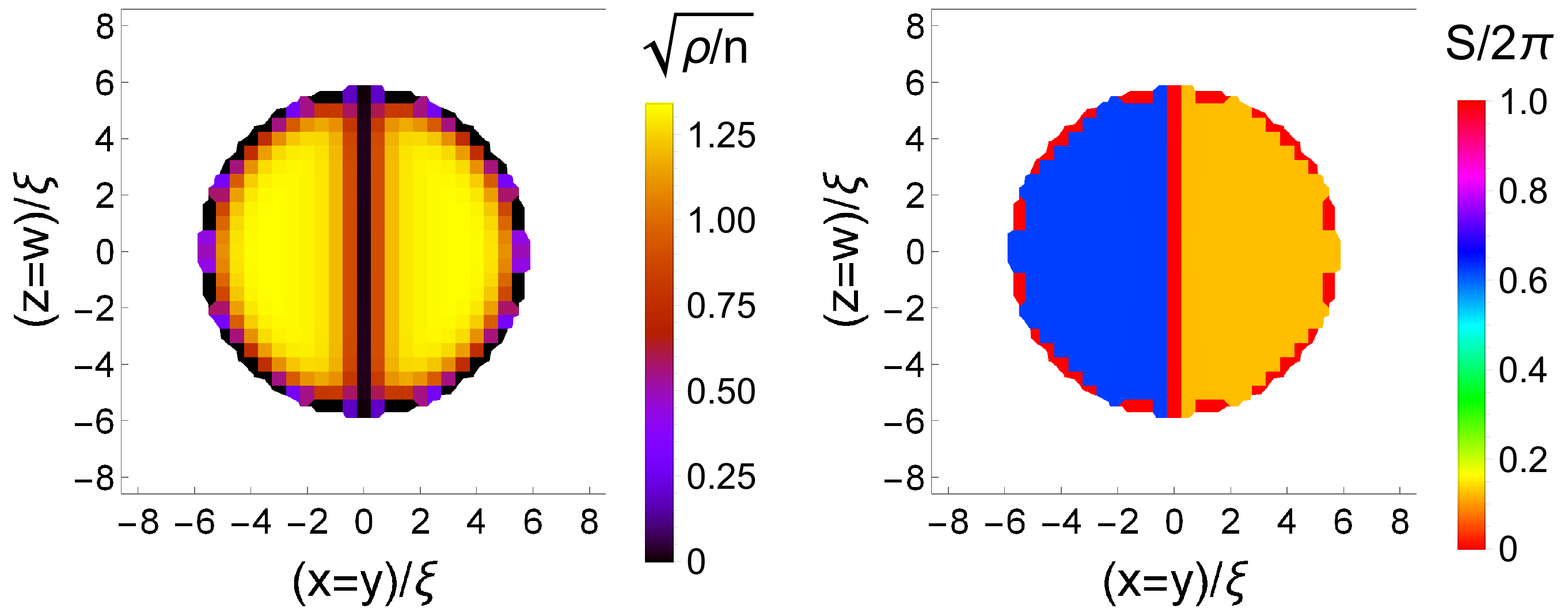}};
                
                \node at ($(letter) + (2*\dx,0)$) {(b)};
                
                \node at ($(letter) + (4*\dx,0)$) {(c)};
                \node at (4*\dx,0) {\includegraphics[width=\picwidth]{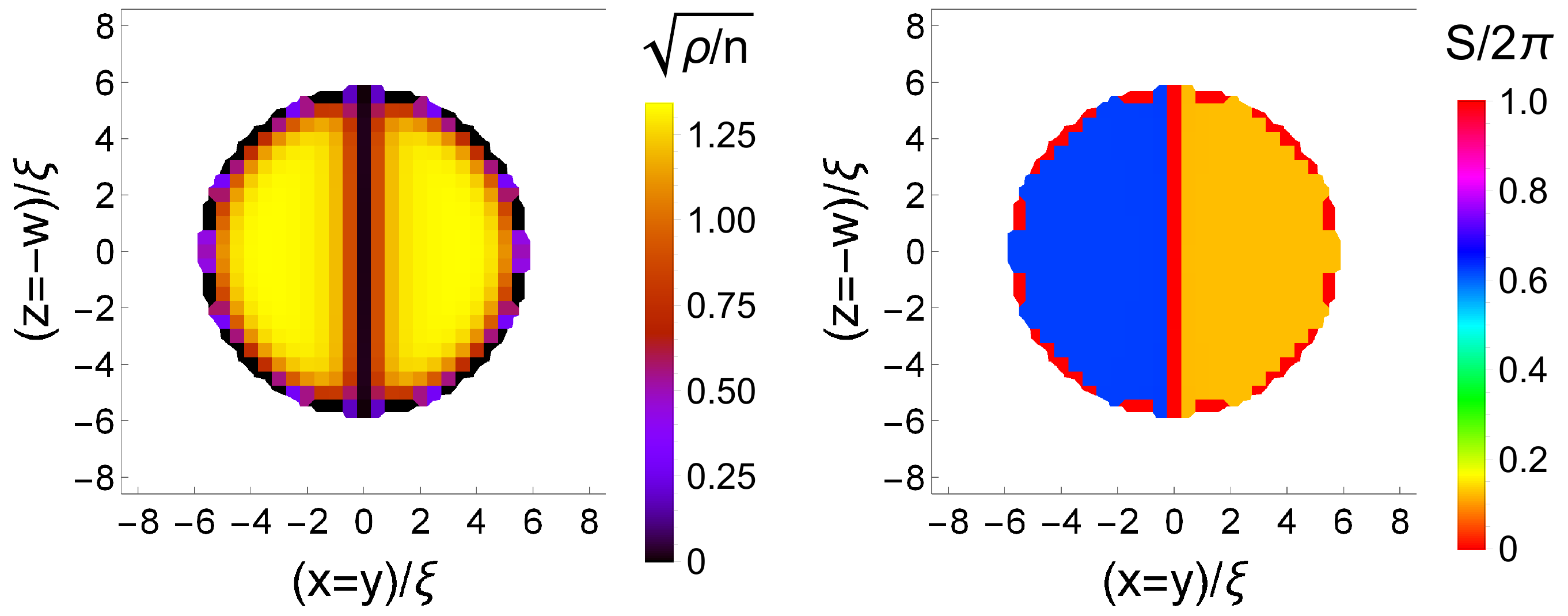}};
                
                \node at ($(letter) + (6*\dx,0)$) {(d)};
                
                \node at ($(letter) + (0,-2*\dy)$) {(e)};
                \node at (0,-2*\dy) {\includegraphics[width=\picwidth]{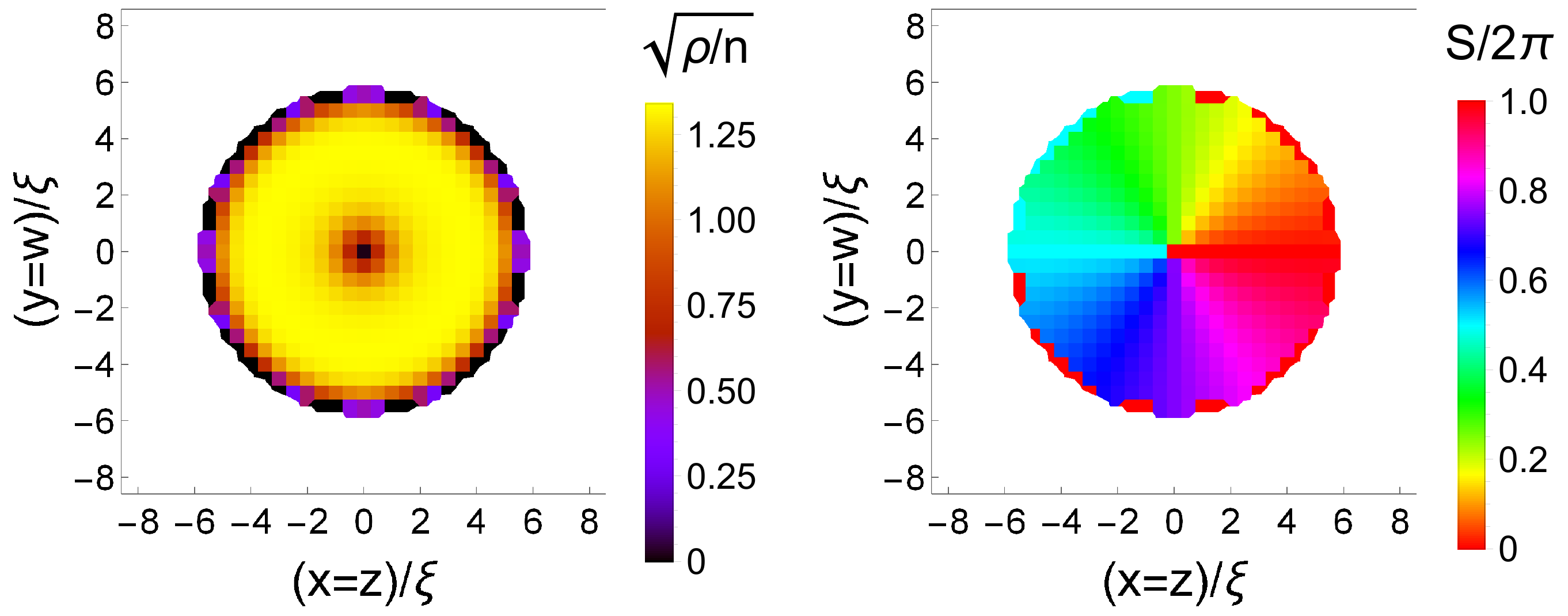}};
                
                \node at ($(letter) + (2*\dx,-2*\dy)$) {(f)};
                
                \node at ($(letter) + (4*\dx,-2*\dy)$) {(g)};
                \node at (4*\dx,-2*\dy) {\includegraphics[width=\picwidth]{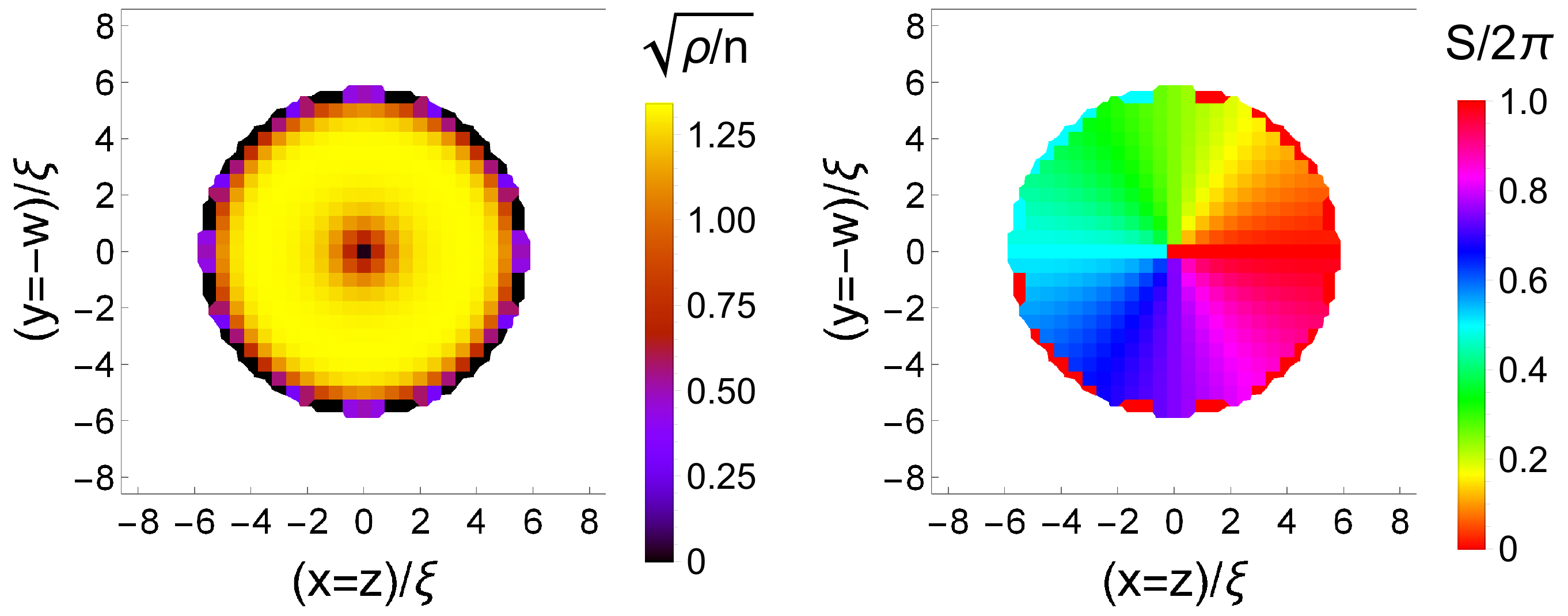}};
                
                \node at ($(letter) + (6*\dx,-2*\dy)$) {(h)};
            \end{tikzpicture}
            \caption{Density (a,c,e,g), and phase (b,d,f,h) profiles for 2D cuts of the numerical stationary state under simple rotation. These cuts are given by (a,b) \(x=y\) and \(z=w\), (c,d) \(x=y\) and \(z=-w\), (e,f) \(x=z\) and \(y=w\), (g,h) \(x=z\) and \(y=-w\). Numerical calculations were performed for a superfluid confined within a discretized 4D hyper-sphere of radius \(8.25 \xi\) and resolution \(0.5\xi\); this discretization is reflected in the pixelation, particularly at the boundaries of the plots. The observed density and phase profiles are in good agreement with a single vortex plane [Eq~\ref{eq:single}].}
            \label{fig:Simple2DCuts}
    \end{figure*}

    We have verified this minimal vortex structure numerically by performing imaginary time evolution on the full 4D GPE under simple rotation in the plane orthogonal to the expected vortex core (i.e. [Eq~\eqref{eq:GPER}] with \(\Omega_1\neq0, \Omega_2=0\)). The corresponding density and phase profiles for the numerical stationary state are shown for selected 2D cuts in Fig~\ref{fig:Simple2DCuts}. Here, the initial state was chosen as detailed in Section II and the rotation frequency was chosen as \(2\Omega_{\text{crit}}^{2D}\). These numerical calculations were performed within a discretized 4D hyper-sphere of radius \(8.25 \xi\), and with resolution \(0.5\xi\).
    
    As can be seen in Fig~\ref{fig:Simple2DCuts}, the observed density and phase profiles are in good agreement with the single vortex plane [Eq~\ref{eq:single}]. In particular, the density is depleted for the plane defined by \(z=0\) and \(w=0\), as is expected for a single vortex plane that approximately takes the form \(\psi\propto (x+iy)\) near the vortex core. Depending on the 2D cut, this vortex plane either appears as a point [see (e) and (g)], as a line [see (a) and (c)] or as a plane [not shown]. Furthermore, around the vortex plane, the superfluid rotates, as can be seen from the winding of the phase in panels (f) and (h) and from the phase jumps in (b) and (d).
    \vspace{-0.5em}
    \subsection{Double Rotations}
    \label{App:Double}
        
\begin{figure*}
            \centering
            \begin{tikzpicture}
                \setlength{\picwidth}{0.49\linewidth}
                \setlength{\dx}{0.25\picwidth}
                \def\dy{1.85};
                \coordinate (letter) at (-1.5*\dx,.75*\dy);
                \node at (letter) {(a)};
                \node at (0,0) {\includegraphics[width=\picwidth]{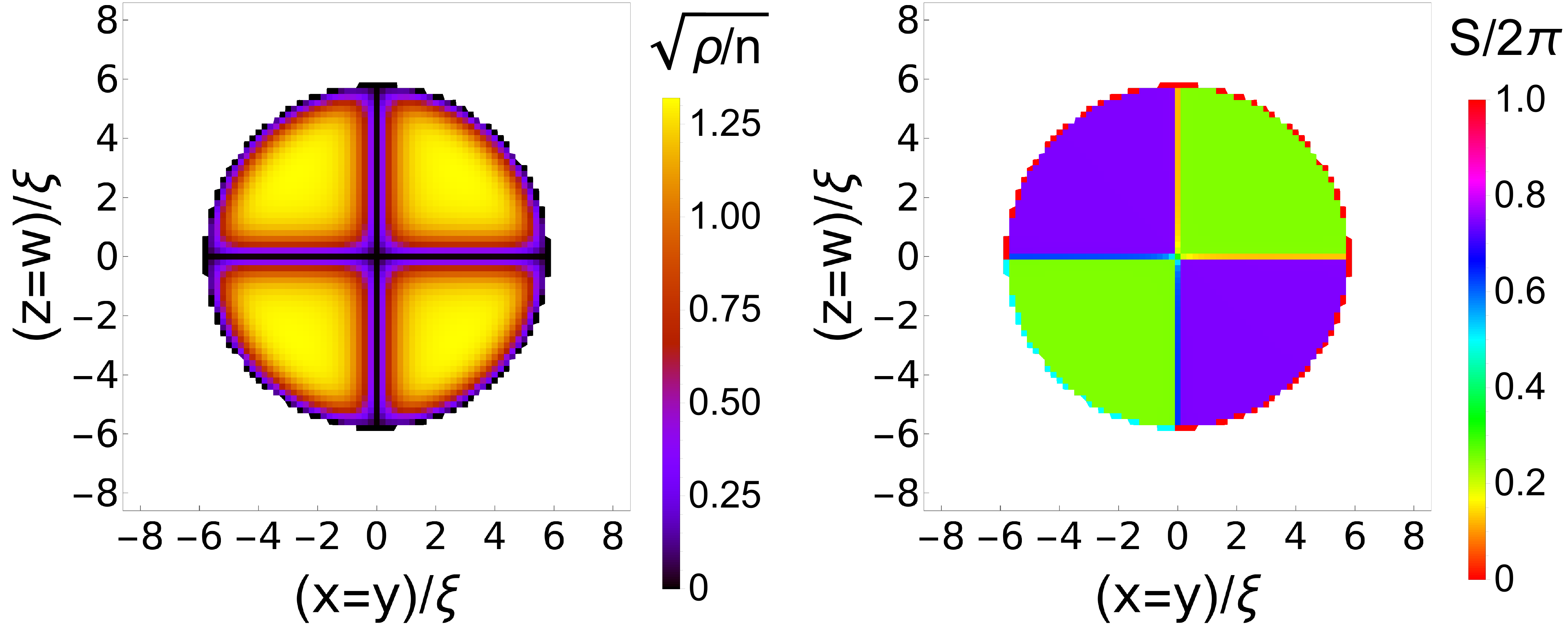}};
                
                \node at ($(letter) + (2*\dx,0)$) {(b)};
                
                \node at ($(letter) + (4*\dx,0)$) {(c)};
                \node at (4*\dx,0) {\includegraphics[width=\picwidth]{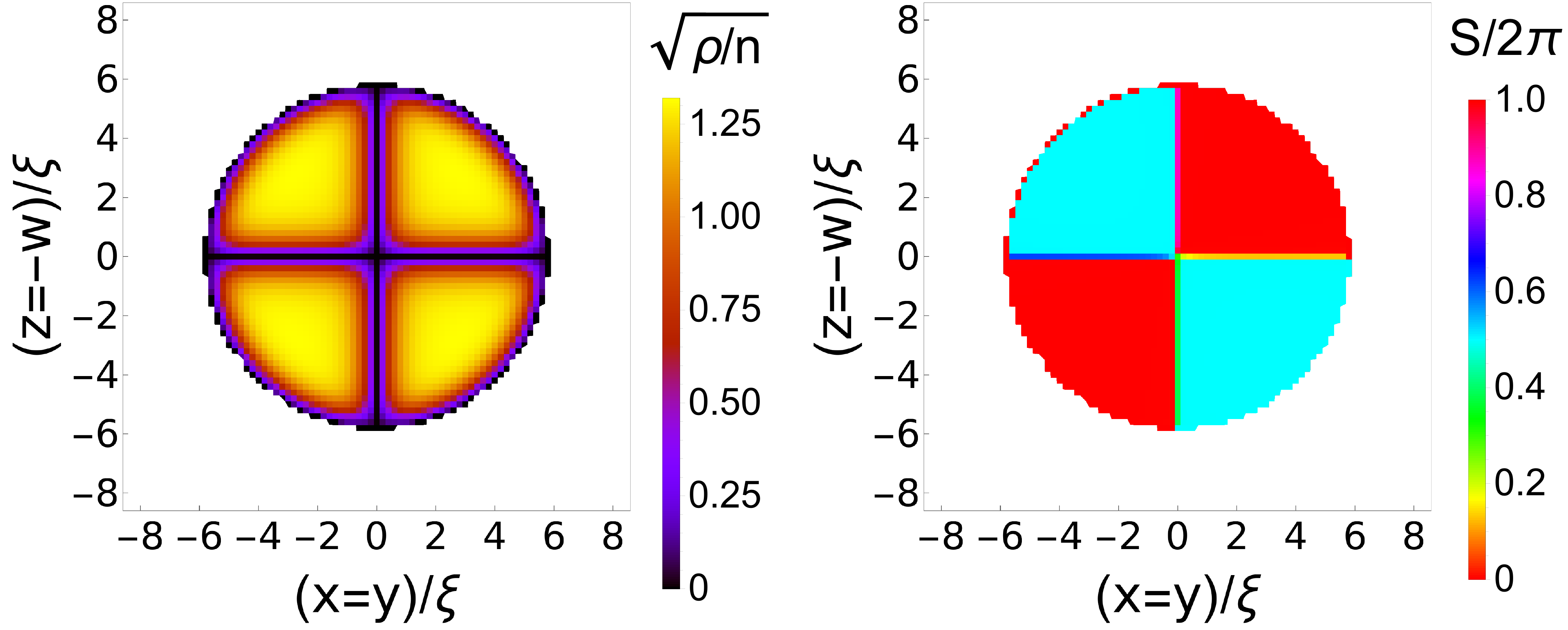}};
                
                \node at ($(letter) + (6*\dx,0)$) {(d)};
                
                \node at ($(letter) + (0,-2*\dy)$) {(e)};
                \node at (0,-2*\dy) {\includegraphics[width=\picwidth]{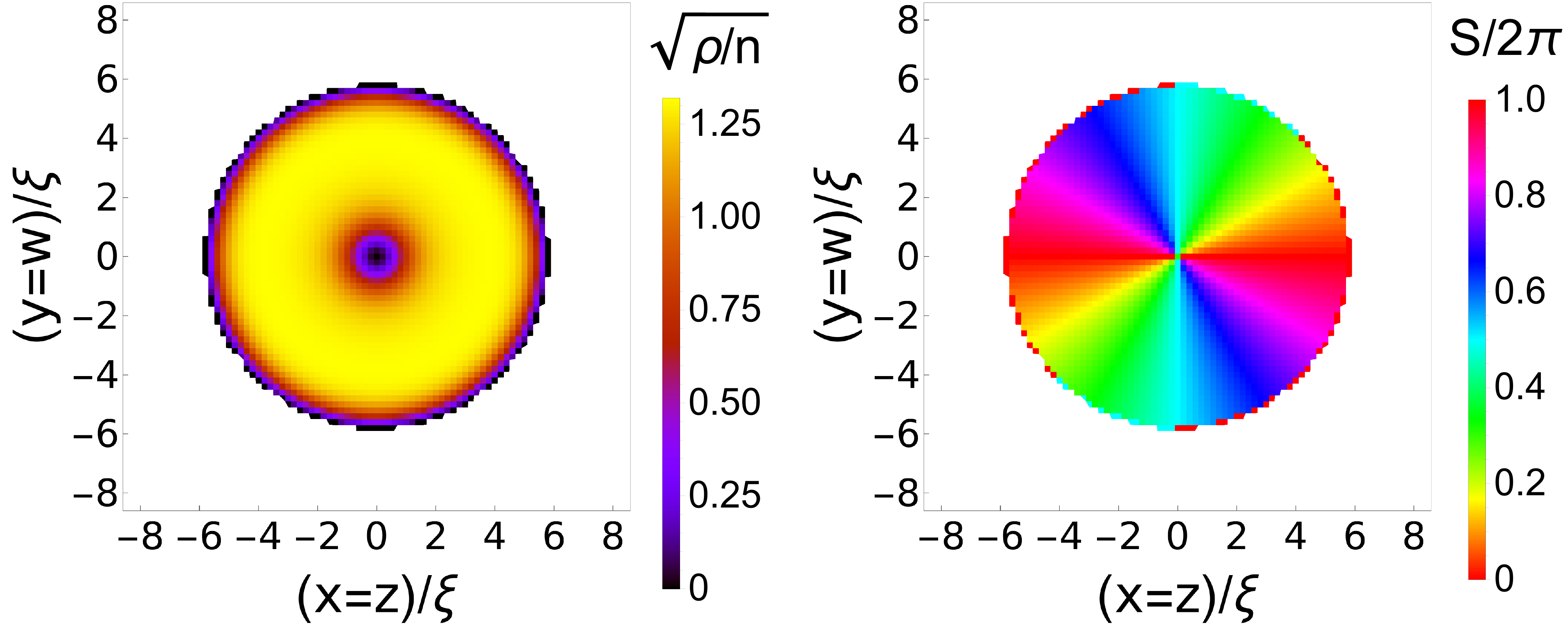}};
                
                \node at ($(letter) + (2*\dx,-2*\dy)$) {(f)};
                
                \node at ($(letter) + (4*\dx,-2*\dy)$) {(g)};
                \node at (4*\dx,-2*\dy) {\includegraphics[width=\picwidth]{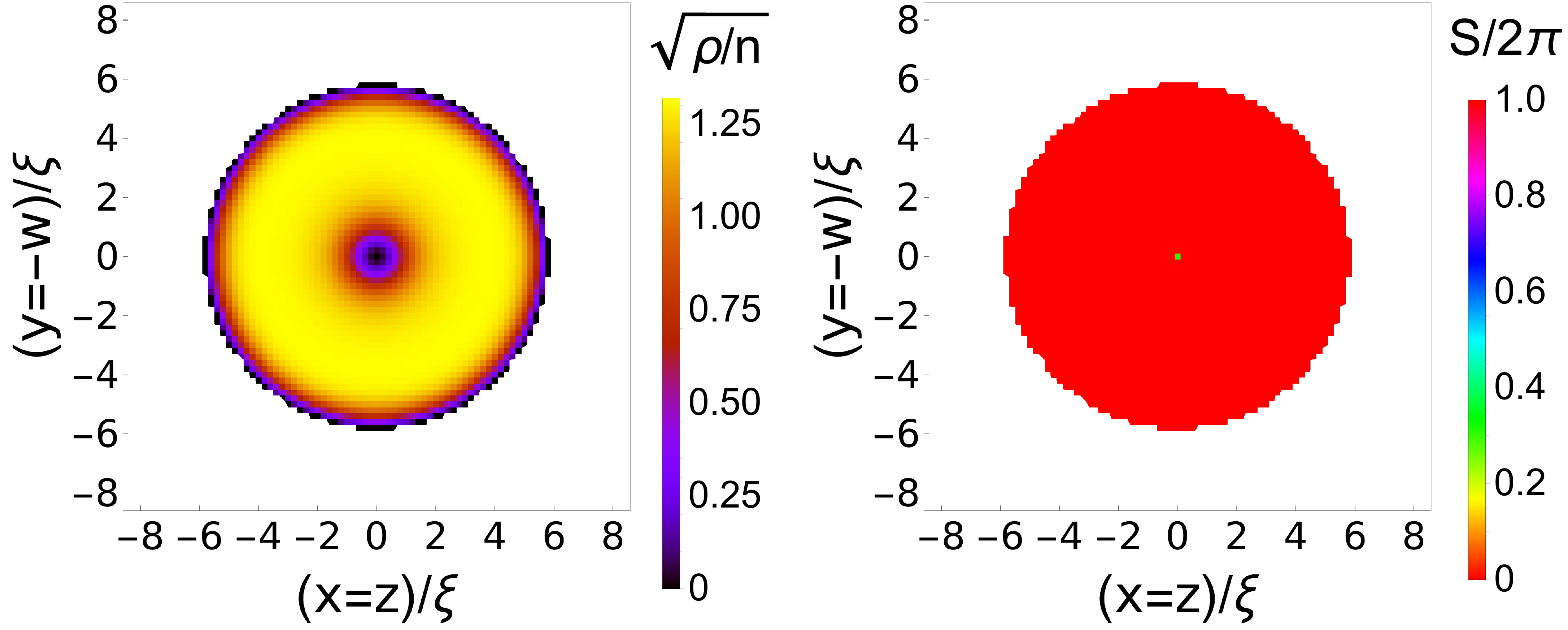}};
                
                \node at ($(letter) + (6*\dx,-2*\dy)$) {(h)};
            \end{tikzpicture}
            \caption{Additional density (a,c,e,g), and phase (b,d,f,h) profiles along 2D cuts of the numerical stationary state studied in Fig~\ref{fig:4DNumerics}. These cuts are given by (a,b) \(x=y\) and \(z=w\), (c,d) \(x=y\) and \(z=-w\), (e,f) \(x=z\) and \(y=w\), (g,h) \(x=z\) and \(y=-w\). The parameters and discretization are detailed in Sec.~\ref{sec:VortexPlanes}. This discretization is reflected in the pixelation, particularly at the boundaries of the plots. The observed density and phase profiles are in good agreement with our numerical ansatz [Eq~\eqref{eq:DoubleAnzatz}], which approximately takes the form \(\psi\propto (x+iy)(z+iw)\) near the vortex cores.}
            \label{fig:2DCuts}
        \end{figure*}

    As we have shown, the double rotation of a 4D superfluid can stabilise a new type of vortex configuration consisting of two vortex planes intersecting at a point. In Fig~\ref{fig:2DCuts}, we plot the density and phase profiles for additional 2D cuts of the numerical stationary state presented in Fig~\ref{fig:4DNumerics}. As can be seen, these profiles have a much richer structure as compared to the case of a single vortex plane shown in Fig~\ref{fig:Simple2DCuts}, as the phase winds simultaneously around both vortex cores with two independent winding numbers. This is also in contrast to 3D systems where two vortex lines may intersect and reconnect over time, but a pair of intersecting vortices is not stabilised by rotation as a stationary state of the system. 
    
    \vspace{-0.5em}
    \subsection{Cuts of the radial profile}
\begin{figure*}
        \centering
        \begin{tikzpicture}
            \setlength{\picwidth}{0.35\linewidth}
            \node (picA) at (-0.25\linewidth,0) {\includegraphics[width=\picwidth]{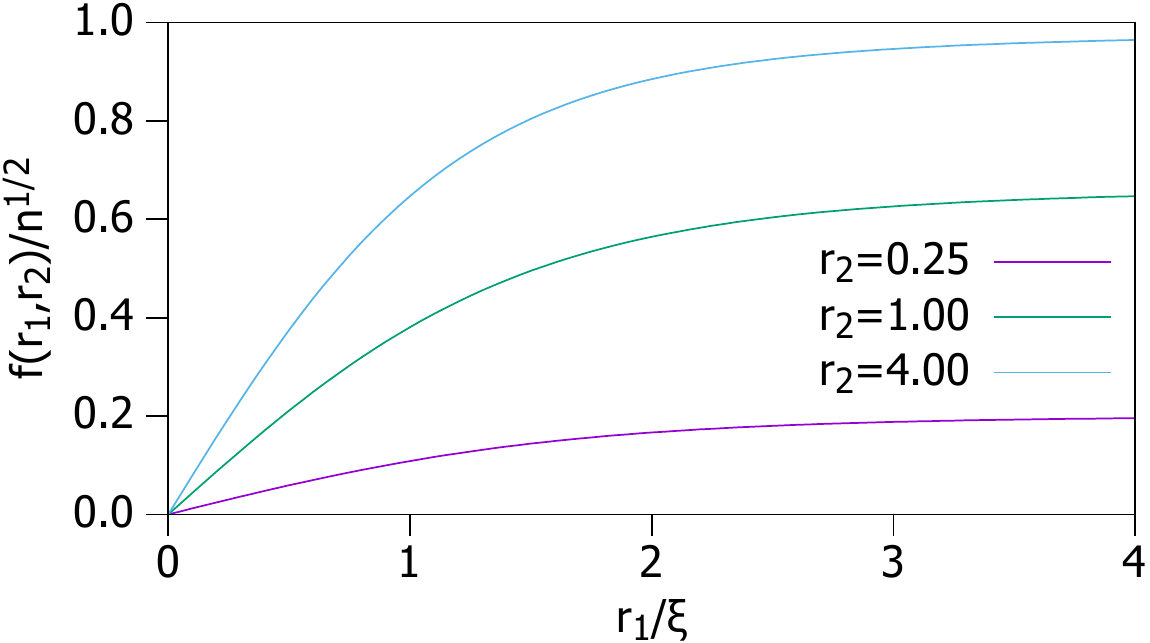}};
            \node (a) at ($(picA)+(-0.5\picwidth,0.25\picwidth)$) {(a)};
            \node (picB) at (0.25\linewidth,0) {\includegraphics[width=\picwidth]{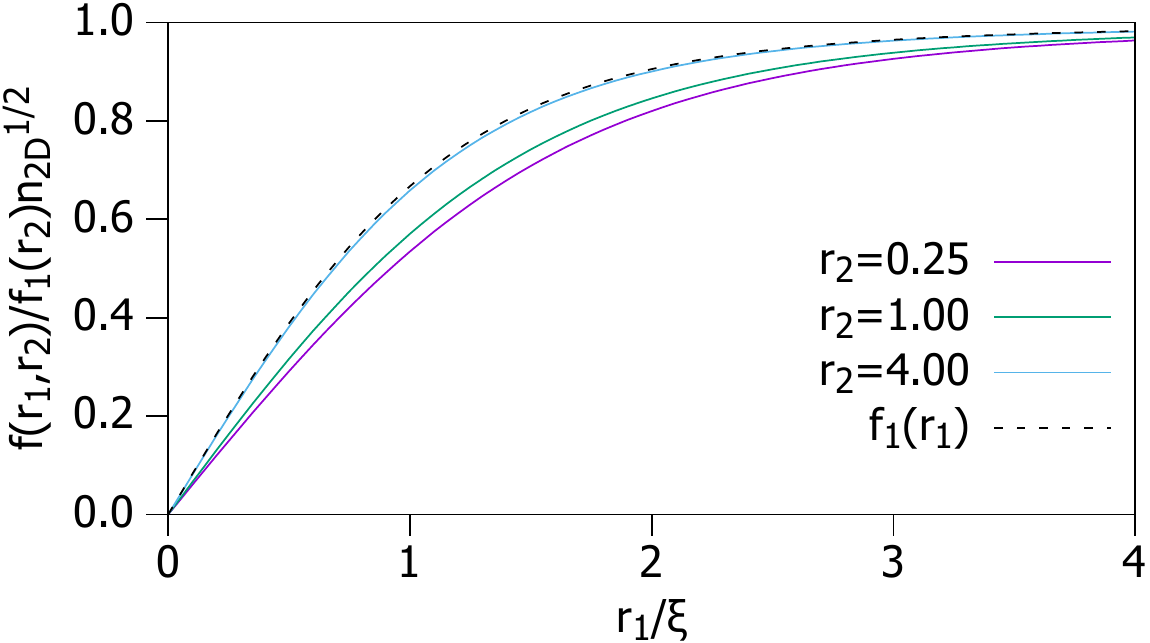}};
            \node (b) at ($(picB)+(-0.5\picwidth,0.25\picwidth)$) {(b)};
        \end{tikzpicture}
        \caption{ (a) Cuts of Fig~\ref{fig:radial} (a), given by fixed values of \(r_2\). (b) As in (a) but rescaled by the 2D vortex density profile \(f_1(r_2)\); note the convergence to \(f_1(r_1)\) for large values of \(r_2\), showing that the order parameter can be well approximated as a product of 2D vortex profiles away from the intersection point. Close to the intersection point, this approximation breaks down, as can be seen from the deviation between these rescaled cuts in this region.}
        \label{fig:RadialCuts}
    \end{figure*}

    As discussed in \red{Section~\ref{sec:VortexPlanes}} and shown in Fig~\ref{fig:radial} (b), we have numerically verified for the solution of the radial equation [Eq~\eqref{eq:f}] that far from the intersection point of the vortex planes the corresponding density profile is well approximated by a product state of the 2D vortex profiles. To visualise this in an alternative way, we have plotted in Fig~\ref{fig:RadialCuts} (a) cuts of Fig~\ref{fig:radial} (a) for specific values of \(r_2\), and then rescaled these by \(f_1(r_2)\) in Fig~\ref{fig:RadialCuts} (b). As shown the rescaled curves approach \(f_1(r_1)\) for large values of \(r_2\), verifying the approximation as expected.
    
    \vspace{-0.5em}
    \subsection{Energy calculation for two intersecting vortex planes in a 4D superfluid}
    \label{App:Energy}
    
    Here, we numerically verify [Eq~\eqref{eq:EnergySum}], which predicts that the energy cost of two intersecting and completely orthogonal vortex planes in a 4D superfluid can be decomposed as a sum of the individual kinetic energies associated with each vortex plane in isolation.
    
    Firstly, we used the numerical solution of the 4D radial density profile presented in Fig~\ref{fig:radial} to calculate the energy of the intersecting vortex planes as a function of system size in each plane. We then produced a fit of this energy to the functional form of [Eq~\eqref{eq:EnergySum}], with the coefficient of \(R_j/\xi\) inside the logarithm as the fitting parameter. From this we obtained \(2.06\) which is very close to the known coefficient of \(2.07\) (in our units) within the logarithmic form of the vortex energy in \(2D\) and \(3D\)~\cite{pethick2002}. This shows that the energy of our numerical solution for the radial equation is consistent with being a sum of two individual vortex energies. 
    
\begin{figure*}
        \centering
        \begin{tikzpicture}
            \setlength{\picwidth}{0.35\linewidth}
            \node (picA) at (-0.25\linewidth,0) {\includegraphics[width=\picwidth]{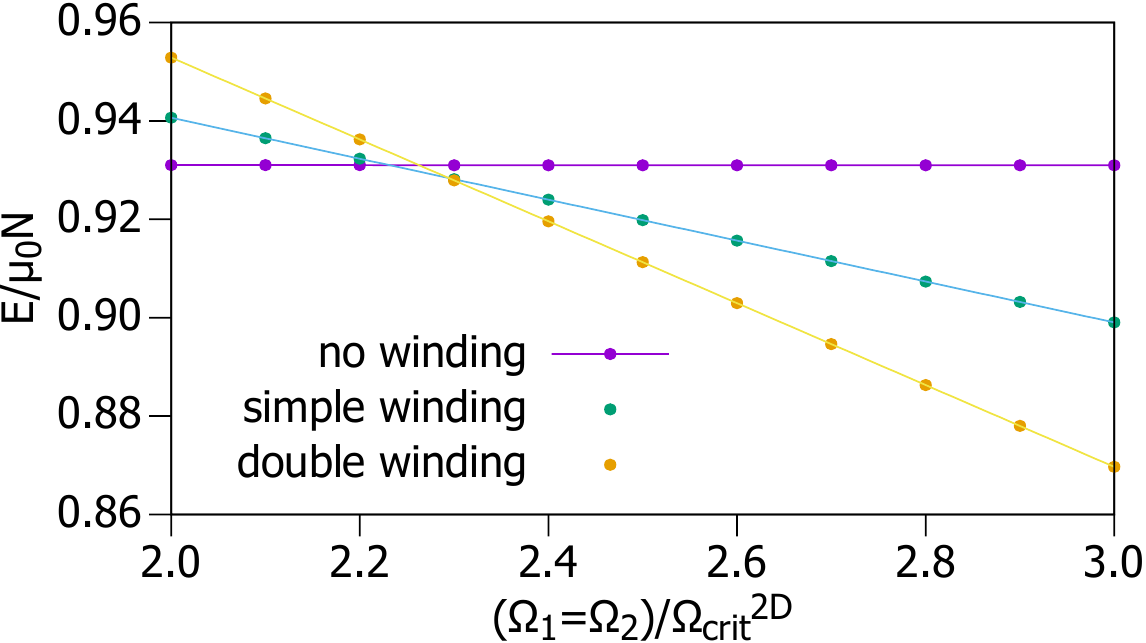}};
            \node (a) at ($(picA)+(-0.5\picwidth,0.25\picwidth)$) {(a)};
            \node (picB) at (0.25\linewidth,0) {\includegraphics[width=\picwidth]{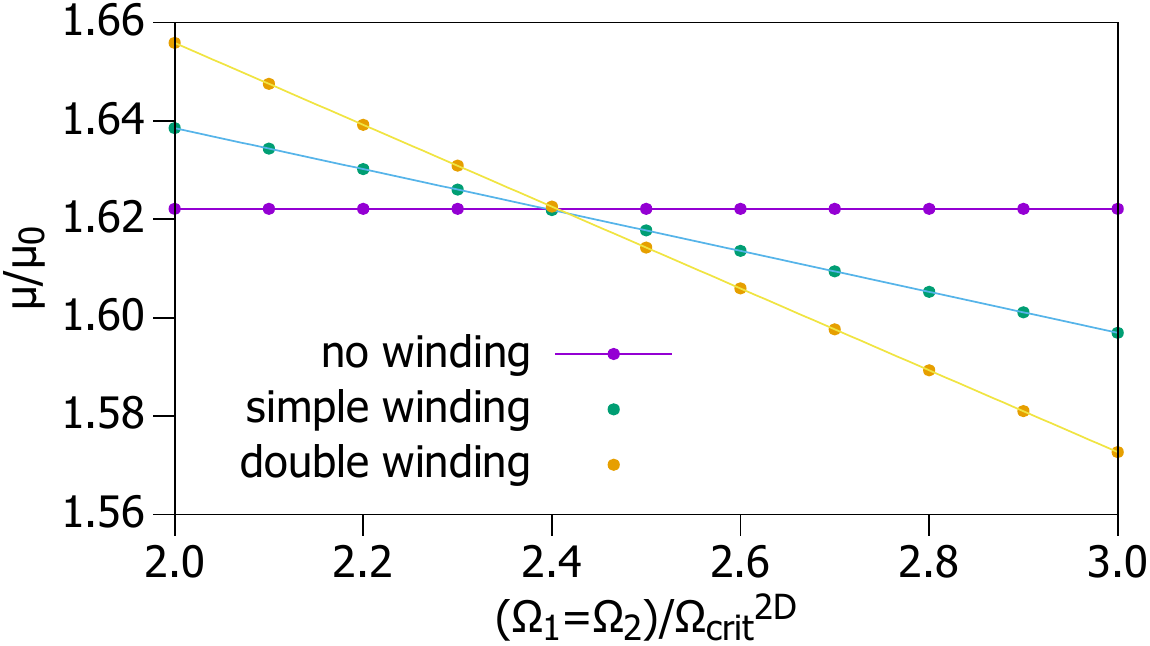}};
            \node (b) at ($(picB)+(-0.5\picwidth,0.25\picwidth)$) {(b)};
        \end{tikzpicture}
        \caption{(a) Energy and (b) Chemical potential of numerical steady states of the 4D doubly rotating GPE [Eq~\eqref{eq:GPER}] with different initial phase profiles. The lines correspond to fits and guides to the eye, respectively, as detailed in the text. The gradient and intercept of these lines give the angular momentum and energy at zero frequency, respectively, of each state, which agree with expected behaviour.}
        \label{fig:energies}
    \end{figure*}

    Secondly, we performed further simulations on a Cartesian 4D grid, with the same parameters as Fig~\ref{fig:4DNumerics}, except for the convergence accuracy which was chosen to be \(10^{-10}\) to speed up calculations. We repeated these calculations for different values of \(\Omega\equiv\Omega_{1}=\Omega_{2}\), ranging between two and three times \(\Omega_{\text{crit}}^{2D}\), in order to numerically verify the expected dependence of the energy on the rotation frequency. Here we used three different initial states: one with no phase winding, one with "simple" winding in one plane, and one with "double" winding in two planes. The resulting values for \(E\) and \(\mu\) as a function of \(\Omega\) are shown in Fig~\ref{fig:energies}, given in units of \(\mu_0\) (the chemical potential of a homogeneous state with no vortices or hard-walls but the same number of particles). We obtain straight lines for each of these data series, showing that each state has well defined angular momentum. 
    
    For the case with no phase winding, we find that \(E/\mu_0 N = 0.931\) and \(\mu /\mu_0= 1.622\) are constants which do not depend on frequency, as expected; this data series is therefore plotted with a straight line joining the dots as a guide to the eye. For the double winding case, we have performed a linear fit, obtaining \(E/\mu_0 N = 1.119 - 0.083\Omega/\Omega_{\text{crit}}^{2D}\) and \(\mu/\mu_0 = 1.822 - 0.083\Omega/\Omega_{\text{crit}}^{2D}\). The gradient, \(-0.083\), is equal to \(-2\Omega_{\text{crit}}^{2D}/\mu_0\), meaning that this is the expected gradient of \(-2\) corresponding to particles having one unit of angular momentum in each plane of rotation. For the simple winding case, we fix the gradient to be half that of the double winding line, since this state has angular momentum in only one of the two planes, and perform a linear fit with only the y intercept as a free parameter. We then obtain \(E/\mu_0 N = 1.023\) and \(\mu/\mu_0=1.722\) when \(\Omega=0\). This gives an energy cost of \(0.188=1.119-0.931\) for the intersecting vortex planes and \(0.092=1.023-0.931\) for the single plane, as compared to the state with no vortices. We expect from [Eq~\eqref{eq:EnergySum}] that these energy costs should be related by a simple factor of two for this geometry, and indeed we find numerically that \(0.188 - 2\times0.092 \simeq 0\).
    
    \section{Homotopy Theory for 4D Vortex Planes}
    \label{App:Homotopy}
    
    Topological excitations, such as vortices, are characterised by topological invariants through homotopy theory. In this approach, the set of allowed topological charges for a given topological defect is given by the set of homotopy classes of maps from a region enclosing the defect to the order parameter manifold. Furthermore, the associated group structure of this set determines the rules for combining two such defects into one. 
    
    In 4D, a plane is enclosed by a circle, just like a line in 3D, or a point in 2D, such that the corresponding homotopy group (for a complex order parameter) is \(\pi_1(S^1)=\mathbb{Z}\). This group is the same as for vortices in lower dimensions, and tells us that each vortex has an integer winding number, and that when two vortices combine their winding numbers combine additively. For the case of two intersecting vortex planes the enclosing region is a 2D torus, such as the product of a circle in the \(xy\) plane and another circle in the \(zw\) plane. The corresponding homotopy group is therefore given by the set of homotopy classes of maps from \(S^1\times S^1\) to \(S^1\), which is isomorphic to \(\mathbb{Z}\times\mathbb{Z}\)~\cite{hatcher}. This simply means that each vortex plane has its own winding number, and the two are independent, as expected for two vortices.
    
    Note that this topological classification is the same as for a pair of linked vortex lines in 3D, which can also be enclosed by a torus. The configuration of 4D intersecting planes therefore offers a simple way to realise the homotopy classification of linked vortex lines within the ground state of a simple 4D GPE model. In the future, it would be interesting to generalise this model to more complicated order parameters, such as those realised in the various phases of spinor BECs~\cite{Kawaguchi}, as then the homotopy group would gain a richer structure, as has been studied in the context of linked line defects in liquid crystals~\cite{Machon}.

\end{appendices}
 
\bibliographystyle{apsrev4-2}
\bibliography{Bibliography}
	
\end{document}